\documentclass[prb,aps,showpacs,twocolumn]{revtex4}
\usepackage{graphicx,dcolumn,amsmath,color,longtable}

\begin{document}

\title{Atomistic spin-model based on a new spin-cluster expansion technique: Application to the IrMn$_3$/Co interface}
\author{L.\ Szunyogh$^1$}\email{szunyogh@phy.bme.hu}
\author{L.\ Udvardi$^1$}
\author{J.\ Jackson$^{2,3}$}
\author{U.\ Nowak$^2$}
\author{R.\ Chantrell$^3$}

\affiliation{$^1$ Department of Theoretical Physics, Budapest University of Technology and Economics, Budafoki \'ut 8. H1111 Budapest, Hungary}
\affiliation{$^2$ Fachbereich Physik, Universit\"at Konstanz, 78457 Konstanz, Germany}
\affiliation{$^3$ Department of Physics, University of York, York YO10 5DD, United Kingdom}
\date{\today}

\begin{abstract}
In order to derive tensorial exchange interactions and local magnetic anisotropies in itinerant magnetic systems, an approach combining the Spin-Cluster Expansion with the Relativistic Disordered Local Moment scheme is introduced.  The theoretical background and computational aspects of the method are described in detail.   The exchange interactions and site resolved anisotropy contributions for the IrMn$_3$/Co(111) interface, a prototype for an exchange bias system, are calculated including a large number of magnetic sites from both the antiferromagnet and ferromagnet.  Our calculations reveal that the coupling between the two subsystems is fairly limited to the vicinity of the interface. The magnetic anisotropy of the interface system is discussed, including effects of the Dzyaloshinskii-Moriya interactions that appear due to symmetry breaking 
at the interface.
\end{abstract}


\maketitle

\section{Introduction}
Spin models form an important tool in the understanding of magnetic systems, particularly of multilayers systems such as bilayers consisting of ferromagnetic and antiferromagnetic materials, because such models allow the investigation of magnetic configurations and dynamics at the atomic scale.  Bilayers of ferromagnetic and antiferromagnetic materials are important parts of magnetoresistive sensors, used, e.g., in hard disc read heads.  The role of the antiferromagnetic component is to pin the ferromagnetic layer by the so-called exchange bias effect, \cite{meiklejohn,noguesJMMM99} leading to a shift of the hysteresis loop of the ferromagnet.  A number of calculations based on spin models have revealed that the anisotropy of the antiferromagnet and the strength of the interface coupling between antiferromagnet and ferromagnet are crucial for the strength and stability of exchange bias \cite{malozemoffPRB87,miltenyiPRL00,nowakJMMM02,scholtenPRB05} and therefore for the functioning of the device. 

IrMn is the most common antiferromagnet industrially because of the large exchange bias and thermal stability that can be obtained with this material.  Despite the successful application of this material in exchange bias bilayers, there is still no complete microscopic understanding of the most important parameters, the antiferromagnetic anisotropy and the interface coupling.  These properties have recently been inferred by a careful examination of the temperature behaviour of the exchange bias of polygranular thin film bilayers\cite{ogrady2010}.  In order to evaluate the antiferromagnet anisotropy, the magnitude of the exchange bias is fitted to a simple model that depends only upon the grain size distribution, yielding effective anisotropy and coupling parameters.  The origin of the effective anisotropy and interface coupling, due to the microscopic physical properties of the interface, remains however inaccessible to direct experimental study.

On the theoretical side, only few first principles calculations on relativistic effects in Mn-based antiferromagnets have been performed.  Regarding the anisotropy, Umetsu {\em et al.}\cite{umetsuAPL06} calculated the magnetic anisotropy energy (MAE) of L1$_{0}$-type MnTM (TM=Ni, Pd, Pt, Rh and Ir) compounds and revealed that MnIr had the highest MAE with a value of -7.05 meV/unit cell.  Furthermore, a giant second order magnetic anisotropy for L$1_2$ IrMn$_3$ was reported, leading to energy barriers of the order of $K_{\mathrm{eff}}$ = 10.4 meV/unit cell  for rotation of the $T$1 ground state spin-structure around the (111) axis. \cite{IrMn-MAE}  
To the best of our knowledge no first principles calculations of the interface exchange coupling between an antiferromagnet and a ferromagnet have been published to date.
 
The organization of the paper is as follows: in the first part we introduce the concept of the Spin-Cluster Expansion (SCE)  merged with the Relativistic Disordered Local Moment (RDLM) scheme and discuss the details of the first-principles calculations.  Then we present our results for the magnetic moments, the exchange coupling and the magnetic anisotropy in L1$_2$ IrMn$_3$, fcc Co bulk, and for the IrMn$_3$/Co(111) interface. Special attention is paid to the effect of Dzyaloshinskii-Moriya interactions at the interface.

\section{Theory}
\subsection{The Spin-Cluster Expansion}
Mapping of the energy of an itinerant magnetic system onto a spin-Hamiltonian presents a longstanding problem in computational solid state physics.  The most widely used approach relies on an adiabatic decoupling of the longitudinal and transverse spin-fluctuations, determined by the fast motion of mobile electrons,  $\tau_e \simeq 10^{-15}s$, and the slow motion of spins,  $\tau_{sf} \simeq 10^{-13}s$.  Under such circumstances, Time-Dependent Density Functional Theory can be approximated by a Landau-Lifshitz equation of motion for the classical spin variables, i.e., the orientation-field, $\vec{e}(\vec{r},t)$,  as combined with Spin-Density Functional Theory (SDFT)
within the Local Spin Density Approximation (LSDA).~\cite{DLM1}  Further simplification of the theory is represented by the Rigid Spin Approximation, in which the orientation of the spins is allowed to change from atomic cell to atomic cell.  A system of $N$ spin-moments is thus characterized by the set of orientations (spin-configurations) $\left\{  \vec{e}\right\}  =\left\{  \vec{e}_{1},\ldots,\vec{e}_{N}\right\}$, with $|\vec{e}_{i}|=1$.  The energy (grand-potential) of the system, $\Omega(\left\{  \vec{e}\right\})$, determined within LSDA then defines a classical Hamiltonian that can be
used in spin-dynamics simulations.

In order to ensure that the process is tractable, a suitable parametrization must be applied to $\Omega(\left\{  \vec{e}\right\})$.  The simplest reliable approximation, an (extended) Heisenberg model, is of second order:
\begin{equation}
\Omega(\left\{  \vec{e}\right\}) = \Omega_0 + \sum_{i=1}^{N}  \vec{e}_i {\bf K}_i
 \vec{e}_i + \frac{1}{2} \sum_{\substack{i,j=1 \\ (i\ne j)}}^{N}
\vec{e}_i {\bf J}_{ij} \vec{e}_j \; ,
\label{extHeis}
\end{equation}
where ${\bf K}_i$ and ${\bf J}_{ij}$ are the second-order anisotropy matrices (which are traceless and symmetric) and the tensorial exchange interactions, respectively.  The Relativistic Torque Method (RTM), which was introduced in  Refs.~\onlinecite{Rtorque1,Rtorque2}, allows the evaluation of these parameters using relativistic electronic structure calculations.  By using RTM, one is required to choose a reference magnetic configuration; usually the ferromagnetic state is taken.  For complex magnetic structures, however, the use of such a reference state might lead to incorrect results concerning, e.g., the magnetic ground state of the system.~\cite{cr3}  Another drawback of RTM is that for systems with reduced point group symmetry the derivation of the anisotropy matrix, ${\bf K}_i$, might be quite a cumbersome task.  More importantly, RTM is not directly applicable to the calculation of more complicated two-spin interactions, such as bi-quadratic coupling, or multispin interactions, known to be present in magnetic systems at the nano-scale.~\cite{mn1cu111,cr3}

The Spin-Cluster Expansion (SCE) developed by Drautz and F\"ahnle~\cite{SCE1,SCE2} in principle solves the problems mentioned above, in that it provides a systematic parametrization of the adiabatic magnetic energy of general classical spin-systems.  In the following we briefly outline the method.  Considering clusters of  $n$ sites ($0 \le n \le N$), $C_n \subset I_N = \{ 1,2,\dots,N \}$, the functions 
\begin{gather}
\Phi^{L_{i_1},\ldots,L_{i_n}}_{C_n}
\left( \left\{ \vec{e} \right\} \right) =
\frac{1}{(4 \pi)^{(N-n)/2}}
\underset{_{i_k \in C_{n}}}{\prod}Y_{L_{i_k}}
\left( \vec{e}_{i_{k}}\right) \; , \\
L_{i_{k}} = (\ell_{i_{k}}, m_{i_{k}}) \ne (0,0)
\nonumber
\end{gather}
form an orthonormal and complete basis set of the function space over $\left\{ \vec{e} \right\}$.  Here $Y_{L}(\vec{e})$ denote, for convenience, real spherical harmonics.  The grand-potential of the system can, therefore, be readily expanded as
\begin{equation}
\Omega\left(  \left\{  \vec{e}\right\}  \right)  =
\sum_{C_{n}}\sum_{L_{i_{1}},\ldots,L_{i_{n}}\neq\left(  0,0\right)  }
J_{C_{n}}^{L_{i_{1} },\ldots,L_{i_{n}}}  \,
\underset{_{i_{k}\in C_{n}}}{\prod}Y_{L_{i_{k}}}\left(
\vec{e}_{i_{k}}\right) \, ,
\label{SCE}
\end{equation}
where the multispin interactions, $J_{C_{n}}^{L_{i_{1} },\ldots,L_{i_{n}}}$, are defined as
\begin{equation}
J_{C_{n}}^{L_{i_{1}},\ldots,L_{i_{n}}}=
\int d^{2} e_{i_1}\ldots d^{2} e_{i_n}
\left\langle\Omega\right\rangle_{\vec{e}_{i_{1}}\ldots\vec{e}_{i_{n}}}
\underset{_{i_{k}\in C_{n}}}{\prod}Y_{L_{i_{k}}}\left(\vec{e}_{i_{k}}\right).
\label{JC}
\end{equation}
In the above equation, the key quantities of the theory, namely, the restricted averages of the grand-potential are:
\begin{equation}
\left\langle \Omega \right\rangle
_{\vec{e}_{i_1}\ldots \vec{e}_{i_n}}
=\frac{1}{\left(  4\pi\right)^{N-n}}\int d^{2N-2n}%
e_{I_{N}\setminus C_{n}}\,
\Omega \left(  \left\{  \vec {e}\right\}  \right) \, ,
\label{restrave}
\end{equation}
i.e., by fixing the spin-orientation at the sites of cluster $C_{n}$, an average must be taken on all spins outside of the cluster, $i \in I_{N}\setminus C_{n}$.  Restricting ourselves to one-site and two-site interactions, Eq.~(\ref{SCE}) reduces to
\begin{eqnarray}
\Omega \left(  \left\{  \vec{e}\right\}  \right)
& \simeq & \Omega_{0}+\sum_{i}%
\sum_{L\neq\left(  0,0\right)  } J_{i}^{L}\,Y_{L}\left(  \vec{e}%
_{i}\right)  \label{SCE-pair} \\
&+& \frac{1}{2}\sum_{i\neq j}\sum_{L\neq\left(  0,0\right)  }%
\sum_{L^{\prime}\neq\left(  0,0\right)  } J_{ij}^{LL^{\prime}}\,Y_{L}\left(
\vec{e}_{i}\right)  Y_{L^{\prime}}\left(  \vec{e}%
_{j}\right) \, , \nonumber
\end{eqnarray}
with
\begin{equation}
\Omega_{0} = \left\langle \Omega\,\right\rangle \; ,
\label{Omega0}
\end{equation}
\begin{equation}
J_{i}^{L}=\int d^{2} e_{i}\,\left\langle \Omega\,\right\rangle
_{\vec{e}_{i}}Y_{L}\left(  \vec{e}_{i}\right)
\label{Jone}
\end{equation}
and
\begin{equation}
J_{ij}^{LL^{\prime}}=\int d^{2} e_{i}\int d^{2}e_{j}\,
\left\langle \Omega\,\right\rangle _{\vec{e}_{i}\vec
{e}_{j}}\,Y_{L}\left(  \vec{e}_{i}\right)  Y_{L^{\prime}}\left(
\vec{e}_{j}\right) \, .
\label{Jtwo}
\end{equation}
The spin-Hamiltonian, Eq.~(\ref{extHeis}), can easily be related to Eq.~(\ref{SCE-pair}) through the relationships,
\begin{equation}
\vec{e}_{i}\,\mathbf{K}_{i}\,\vec{e}_{i}=\sum_{m=-2}%
^{2}J_{i}^{\left(  2,m\right)  }\,Y_{2,m}\left(  \vec{e}%
_{i}\right)  +\mathrm{const.}
\label{eKe}
\end{equation}
and
\begin{equation}
\vec{e}_{i}\,\mathbf{J}_{ij}\,\vec{e}_{j}=\sum
_{m,m^{\prime}=-1}^{1}
\,J_{ij}^{\left(  1,m\right)  \left(  1,m^{\prime}\right)  }\,
Y_{1,m}\left(  \vec{e}_{i}\right) \,
Y_{1,m^{\prime} }\left(  \vec{e}_{j}\right) \, .
\label{eJe}
\end{equation}

The evaluation of Eq.~(\ref{restrave}) obviously poses a heavy numerical task since it involves integration of the first-principles grand-potential over $N-n$ spin-variables.  In particular, for infinite systems like bulk magnets or magnetic films, it seems extremely demanding to ensure the completion of this task.  A suitable method is therefore needed that can evaluate the restricted averages in a feasible manner.

\subsection{The Relativistic Disordered Local Moment scheme}
The Disordered Local Moment (DLM) scheme~\cite{DLM1} is an extension to conventional SDFT to include transverse spin-fluctuations.  In the DLM scheme the magnetic excitations are modelled by associating local spin-polarization axes with all lattice sites and the orientations $\{\vec{e}\}$ vary very slowly on the time-scale of the electronic motion.  These ``local moment'' degrees of freedom produce local magnetic fields centered at the lattice sites which in turn affect the electronic motions and are self-consistently maintained by them.
The theory of Disordered Local Moments (DLM) in conjunction with the Korringa-Kohn-Rostoker Coherent Potential Approximation (KKR-CPA) was proposed twenty-five years ago by Gy\"orffy et al.~\cite{DLM1} and has recently been generalized by Staunton et al.~\cite{RDLM1,RDLM2,RDLM3} to include relativistic effects.  Here we give a short summary of the method as applied to the consideration of the paramagnetic state.

For spherically symmetric potentials and exchange fields, the local orientation of the spin-magnetization is accounted for by the similarity transformation of the single-site {\it t}-matrix,
\begin{equation}
\underline{t}_{i}\left(  \vec{e}_{i}\right)  =\underline{D}\left(
\vec{e}_{i}\right)  \,\underline{t}_{i}\left(  \vec{e}_z \right)
\underline{D}\left(  \vec{e}_{i}\right)  ^{+}\;,
\label{t-e}
\end{equation}
where for a given energy (not labelled explicitly) $\underline{t}_{i}\left( \vec{e}_z \right)$ stands for the \emph{t}-matrix with exchange field pointing along the local $z$ axis and $\underline{D}\left(  \vec{e}_{i}\right)$ is a (block-wise) projective representation of the SO(3) transformation that rotates the $z$ axis onto $\vec{e}_{i}$.  Note that underlining indicates matrices in the ($\kappa,\mu$) angular-momentum representation.  In the spirit of the single-site coherent-potential approximation (CPA) an orientation independent reference medium is introduced in terms of the effective \emph{t}-matrices, $\underline {t}_{i,c}$, and the corresponding scattering path operator (SPO) matrix,
\begin{equation}
\underline{\underline{\tau}}_{c}=\left(
\underline{\underline{t}}_{c}^{-1}-
\underline{\underline{G}}_{0}\right)  ^{-1}\;.
\label{tau-CPA}
\end{equation}
In the above equation, double underlines denote matrices in site--angular momentum space; $\underline{\underline{t}}_{c}$ is diagonal in site indices while $\underline{\underline{G}}_{0}$ stands for the matrix of structure constants.  The effective \emph{t}-matrices are determined by the condition that the excess scattering matrices,
\begin{equation}
\underline{X}_{i}\left(  \vec{e}_{i}\right)  =\left[  \left(
\underline{t}_{i,c}^{-1}-
\underline{t}_{i}^{-1}\left(  \vec{e}_{i}\right)  \right)
^{-1}-\underline{\tau}_{ii,c}\right]  ^{-1}\;,
\label{Xmatrix}
\end{equation}
vanish when averaged over all possible orientations with an equal probability in the paramagnetic state,
\begin{equation}
\frac{1}{4 \pi} \int\underline{X}_{i}\left(  \vec{e}_{i}\right)
d^2 e_{i}=\underline{0}\;. 
\label{CPA-1}
\end{equation}
The above condition can also be reformulated in terms of the site-diagonal SPO matrices,
\begin{equation}
\left\langle \underline{\tau}_{ii}\left(  \left\{  \vec{e}\right\}
\right)  \right\rangle =
\frac{1}{4\pi}\int\left\langle \underline{\tau}
_{ii}\right\rangle _{\vec{e}_{i}}d^2 e_{i}=\underline{\tau}%
_{ii,c}\;, 
\label{CPA-2}
\end{equation}
where the so-called restricted average of the site-diagonal SPO matrix can be expressed as,
\begin{equation}
\left\langle \vec{\tau}_{ii}\right\rangle _{\vec{e}_{i}}%
=\underline{\tau}_{ii,c}\underline{D}_{i}\left(  \vec{e}_{i}\right)  \;,
\end{equation}
with
\begin{eqnarray}
\underline{D}_{i}\left(  \vec{e}_{i}\right)
& = & \underline{I}+\underline
{X}_{i}\left(  \vec{e}_{i}\right)  \underline{\tau}_{ii,c}
\label{Dmatrix-1} \nonumber \\
& = &
\left[  \underline
{I}+\left(  \underline{t}_{i}^{-1}\left(  \vec{e}_{i}\right)
-\underline{t}_{i,c}^{-1}\right)  \underline{\tau}_{ii,c}\right]  ^{-1}\;.
\label{Dmatrix-2}
\end{eqnarray}

As discussed in the previous section, we limit our consideration by neglecting longitudinal spin-fluctuations; i.e.  we suppose that the LSDA potentials and exchange fields are independent of the orientational state $\left\{ \vec{e} \right\}$.  The magnetic force theorem\cite{Jansen99} then implies that the LSDA total energy can be replaced by the single-particle energies (band energy) and the grand-potential at zero temperature can be expressed as
\begin{align}
\Omega\left( \left\{  \vec{e}\right\}  \right)
&  =E_{LSDA}\left(  \left\{  \vec{e}\right\} \right)
 -\varepsilon_F N\left(  \left\{
\vec{e}\right\}  \right)  \nonumber \\
&  \simeq -\int^{\varepsilon_F} d\varepsilon\,N\left(
\varepsilon;\left\{  \vec{e}\right\}  \right)
\;.
\label{Omega}
\end{align}
Here $\varepsilon_F$ is the Fermi energy and $N\left(  \varepsilon; \left\{  \vec{e}\right\}  \right)$ denotes the integrated density of states.  Using Lloyd's formula\cite{Lloyd67} and taking the coherent medium as reference, Eq.~(\ref{Omega})  can be recast as
\begin{align}
& \Omega\left(\left\{  \vec{e}\right\}  \right)
  = \Omega_{c}-\frac{1}{\pi}\sum_{i}
\operatorname{Im}\int^{\varepsilon_F} d\varepsilon
\,\ln\det\underline{D}_{i}\left( \vec{e}_{i}\right)
  \nonumber \\  &
-\frac{1}{\pi}\sum_{k=1}^{\infty}\frac{1}{k}\sum_{i_{1}\neq i_{2}%
\neq\ldots\neq i_{k}}\,\operatorname{Im}\int^{\varepsilon_F} d\varepsilon \,
\nonumber\\ &
 \mathrm{Tr}\left(
\underline{X}_{i_1}\left( \vec{e}_{i_1}\right)
\underline{\tau}_{c,i_1 i_2}
\underline{X}_{i_2}\left(  \vec{e}_{i_2}\right)
\ldots
\underline{X}_{i_k}\left( \vec{e}_{i_k}\right)
\underline{\tau}_{c,i_k i_1} \right)
\;,  \label{Omega-exp}
\end{align}
with
\begin{equation}
\Omega_{c}=-\int^{\varepsilon_F} d\varepsilon \,N_{0}\left(
\varepsilon\right)  -\frac{1}{\pi}\operatorname{Im}
\int^{\varepsilon_F} d\varepsilon
 \,\ln\det\underline{\underline{\tau}}_{c}\left(
\varepsilon\right)  \;.
\end{equation}
where $N_{0}\left( \varepsilon\right)$ represents the integrated density of states of 
the free electrons.

Eq.~(\ref{Omega-exp}) is, in principle, valid for reference systems fixed by any set of {\it t}-matrices, $\underline{\underline{t}}_{c,i}$.  The particular choice of the RDLM effective medium, defined by the condition Eq.~(\ref{CPA-1}), is very useful in finding a suitable approximation for the restricted averages in Eq.~(\ref{restrave}) and consequently for the spin-cluster interactions in Eq.~(\ref{JC}).  More specifically, when taking averages we neglect the so-called back-scattering contributions in the third term of the right-hand side of Eq.~(\ref{Omega-exp}), i.e., those for which any site outside of the chosen cluster, $i_l \in I_N \setminus C_n \; (l=1,\ldots,k)$, occurs at least twice.  Under this restriction, which is consistent with the single-site CPA,\cite{Butler85} we obtain for the constant term in Eq.~(\ref{Omega0}),
\begin{equation}
\left\langle \Omega\,\right\rangle =
  \Omega_{c}-\frac{1}{\pi}\sum_{i}
\operatorname{Im}\int^{\varepsilon_F} d\varepsilon
\, \frac{1}{4\pi} \int d^2 e_{i}
\,\ln\det\underline{D}_{i}\left( \vec{e}_{i}\right)  \; .
\label{Omega0-b}
\end{equation}
The one-site restricted average of the grand potential can be expressed as,
\begin{align}
\left\langle \Omega\,\right\rangle_{\vec{e}_i} =&  \Omega_{c}
-\frac{1}{\pi} \operatorname{Im}\int^{\varepsilon_F} d\varepsilon
\,\ln\det\underline{D}_{i}\left( \vec{e}_{i}\right) 
\nonumber \\
& -\frac{1}{\pi}\sum_{j (\ne i)}
\operatorname{Im}\int^{\varepsilon_F} d\varepsilon
\, \frac{1}{4\pi} \int d^2 e_{j}
\,\ln\det\underline{D}_{j}\left( \vec{e}_{j}\right)  \; ,
\label{Omega1-b}
\end{align}
from which one obtains for the one-site expansion coefficient,
\begin{equation}
J^{L}_i =
-\frac{1}{\pi} \operatorname{Im}\int^{\varepsilon_F} d\varepsilon
\int d^2 e_{i}
\,\ln\det\underline{D}_{i}\left( \vec{e}_{i}\right) \, Y_L(\vec{e}_i) \; .
\label{Jone-b}
\end{equation}
From the two-site ($i\ne j$) restricted averages,
\begin{align}
& \left\langle \Omega\right\rangle _{\vec{e}_i \vec{e}_j} 
=\Omega_{c} \nonumber \\ &
-\frac{1}{\pi}\operatorname{Im}\int^{\varepsilon_F} d\varepsilon
\,\left(  \ln\det\underline{D}_{i}\left( \vec{e}_{i}\right)  +
          \ln\det\underline{D}_{j}\left( \vec{e}_{j}\right) \right)
\nonumber \\ &
-\frac{1}{\pi}\operatorname{Im} \int^{\varepsilon_F} d\varepsilon
\sum_{l (l \ne i,j)} \frac{1}{4\pi} \int d^2 e_{l}
\,\ln\det\underline{D}_{l}\left( \vec{e}_{l}\right)
\nonumber \\ &
-\frac{1}{\pi}\sum_{k=1}^{\infty}\frac{1}{k}\,
\operatorname{Im} \int^{\varepsilon_F} d\varepsilon  \,
\mathrm{Tr} \left[ \left(
\underline{X}_i\left(\vec{e}_i\right) \underline{\tau}_{c,ij}
\underline{X}_j\left(\vec{e}_j\right) \underline{\tau}_{c,ji} \right)^{k}
\right] \;,
\end{align}
and using Eq.~(\ref{Jtwo}) the following closed expression for the two-spin interactions can be deduced,
\begin{align}
J^{LL'}_{ij} & = -\frac{1}{\pi} \operatorname{Im}\int^{\varepsilon_F} d\varepsilon
\iint d^2 e_{i} \, d^2 e_{j} \, 
\nonumber \\ &
\mathrm{Tr} \ln\left(  \underline{I}-
\underline{X}_i\left(\vec{e}_i\right)  \underline{\tau}_{c,ij}
\underline{X}_j\left(\vec{e}_j\right)  \underline{\tau}_{c,ji}  \right)
\, Y_L(\vec{e}_i) \, Y_{L'}(\vec{e}_j)
\;.
\label{Jtwo-b}
\end{align}
In the Appendix we derive the non-relativistic limit of the above expression.

It is also straightforward to show that a general multispin interaction as defined in Eq.~(\ref{JC}) is given within the RDLM scheme as
\begin{align}
& J_{C_{n}}^{L_{i_{1}},\ldots,L_{i_{n}}}= 
-\frac{1}{\pi} \operatorname{Im}\int^{\varepsilon_F} d\varepsilon \,
\sum_{k \ge n}^{\infty}\frac{1}{k} \sum_{i_{1}\neq i_{2}\neq\ldots\neq i_{k}}
\nonumber \\
& 
\qquad \qquad \quad  \idotsint d^{2} e_{i_1}\ldots d^{2} e_{i_n} \,  \prod_{l=1}^n
Y_{L_{i_{l}}}\left(\vec{e}_{i_l}\right)
\nonumber \\ &
 \mathrm{Tr}\left(
\underline{X}_{i_1}\left( \vec{e}_{i_1}\right)
\underline{\tau}_{c,i_1 i_2}
\underline{X}_{i_2}\left(  \vec{e}_{i_2}\right)
\ldots
\underline{X}_{i_k}\left( \vec{e}_{i_k}\right)
\underline{\tau}_{c,i_k i_1} \right) \,
 \, ,
\label{JC-b}
\end{align}
where each term on the right-hand side must contain all the sites in cluster $C_n$.

\section{Application to
the IrMn$_3$/Co(111) interface}
\subsection{Computational details}
We performed self-consistent calculations for the ordered bulk L1$_2$ IrMn$_3$ alloy, for fcc bulk Co and for the IrMn$_3$/Co(111) interface in terms of the Screened Korringa-Kohn-Rostoker (SKKR) method \cite{skkr1,skkr2}.  Because the magnetic structure of the IrMn$_3$/Co(111) interface was not known a priori, we employed in these calculations the scalar-relativistic DLM approach that is representative of the paramagnetic state of the system.  Here we solved the CPA condition, Eq. (\ref{CPA-nr}), with a relative accuracy of $10^{-5}$ for the effective {\it t}-matrices.  The local spin--density approximation as parametrized by Vosko et al.\cite{voskoCJP80} was applied, the effective potentials and fields were treated within the atomic sphere approximation with an angular momentum cut--off of $\ell_{max}=2$.  The energy integrations were performed by sampling 12 points on a semi-circular path in the upper complex semi-plane and 84 points were selected in the 2D Brillouin zone, i.e., in one fourth of the fcc(111) surface Brillouin zone, for the necessary {\it k}-integrations.  In all cases we used a parent fcc lattice structure with the lattice constant of $L$1$_2$ IrMn$_3$, $a=3.785$ \AA~\cite{tomenoJAP99,sakumaPRB03}, with no attempt at geometric relaxation.

To ensure that both the IrMn$_3$ and Co components shared the same two-dimensional translational periodicity, which is necessary within the layered SKKR method for an interface, the calculations for Co were carried out with four atoms per cubic unit cell according to the fcc lattice structure.  The interface calculations were performed for a symmetric system consisting of two 6 monolayer (ML) thick IrMn$_3$ layers sandwiching a 12 ML Co layer from both sides.  Considering four atoms in a unit cell for each layer (three Mn and one Ir atoms in each IrMn$_3$ layer and four Co atoms in each Co layer), the total 2D unit cell of the interface system contained 96 atoms.  In addition, to ensure smooth boundary conditions, the entire system was sandwiched between two perfect semi-infinite bulk IrMn$_3$ systems as required by the SKKR method~\cite{SKKR,skkr1}.

In order to evaluate the integrals over the orientations in Eqs. (\ref{Jone-b}) and (\ref{Jtwo-b}) we used the efficient Lebedev-Laikov 
scheme.\cite{Lebedev99} The energy integration in Eq. (\ref{Jone-b})  was performed by introducing a damping through the Fermi
function with $T=50 K$. The corresponding integral can then be transformed into a sum over the fermionic Matsubara poles 
and the real part instead of the imaginary part of the logarithm of complex numbers has to be considered, thereby avoiding the
well-known phase problem associated with Lloyd's formula.\cite{Zeller05}   In Eq. (\ref{Jtwo-b}) the above mentioned contour integration
can safely be used, since the expression, Tr$\, \ln (\underline{I}-\underline{A})$, in
the integrand can be expanded into a rapidly converging power series.  

\subsection{Magnetic moments in the DLM state}
The self-consistent field calculations resulted in a local magnetic moment of 2.20 $\mu_B$ for the Mn atoms in L1$_2$ IrMn$_3$.  This value is about 15~\% less than the value we obtained for the the spin-moment in a relativistic calculation for the ordered triangular ($T1$) state.~\cite{IrMn-MAE}  Such a softening of the magnetic moments is expected in the paramagnetic DLM state due to the vanishing Weiss--field acting on the spins.  This reduction of the local moment is not necessarily accompanied by a softening of the exchange coupling, however, as is demonstrated below.  The calculated local magnetic moment of the fcc bulk Co in the DLM state is 1.62$\mu_B$, in agreement with the experimentally and theoretically obtained values for ferromagnetic bulk Co.

\begin{figure}[htb]
\includegraphics{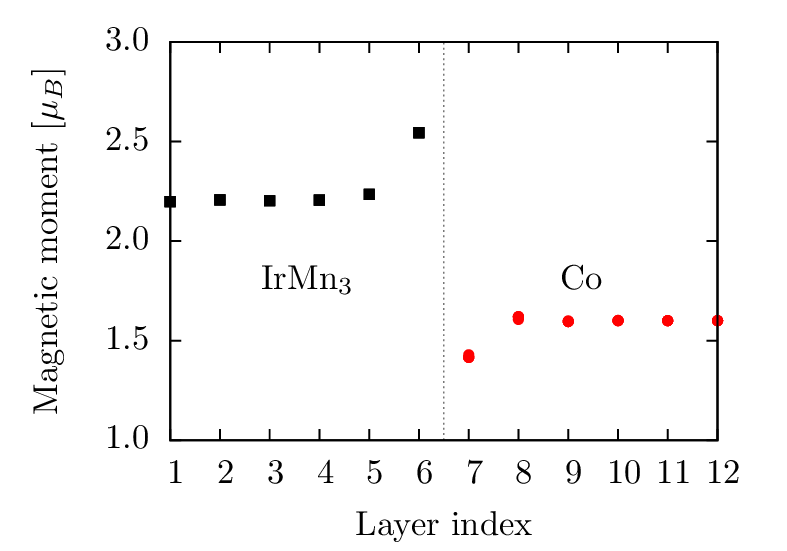}
\caption{(Colour online) Calculated magnetic moments of Mn and Co atoms for the IrMn$_3$/Co(111) interface.  The IrMn$_3$ and Co sides of the interface, marked by the vertical dotted line, correspond to layers labelled by 1 to 6 and 7 to 12, respectively.}
\label{fig:moments}
\end{figure}

In Fig.~\ref{fig:moments} the magnetic moments for the IrMn$_3$/Co(111) interface system are shown as calculated in the DLM state.  Note that only one half of the symmetric system is shown.  Clearly, only moments in close vicinity of the interface are substantially different from their bulk values: the Mn moments increase to 2.54$\mu_B$ and the Co moments reduce to 1.42$\mu_B$.  

\subsection{Isotropic exchange interactions}
\begin{figure}[htb]
\includegraphics{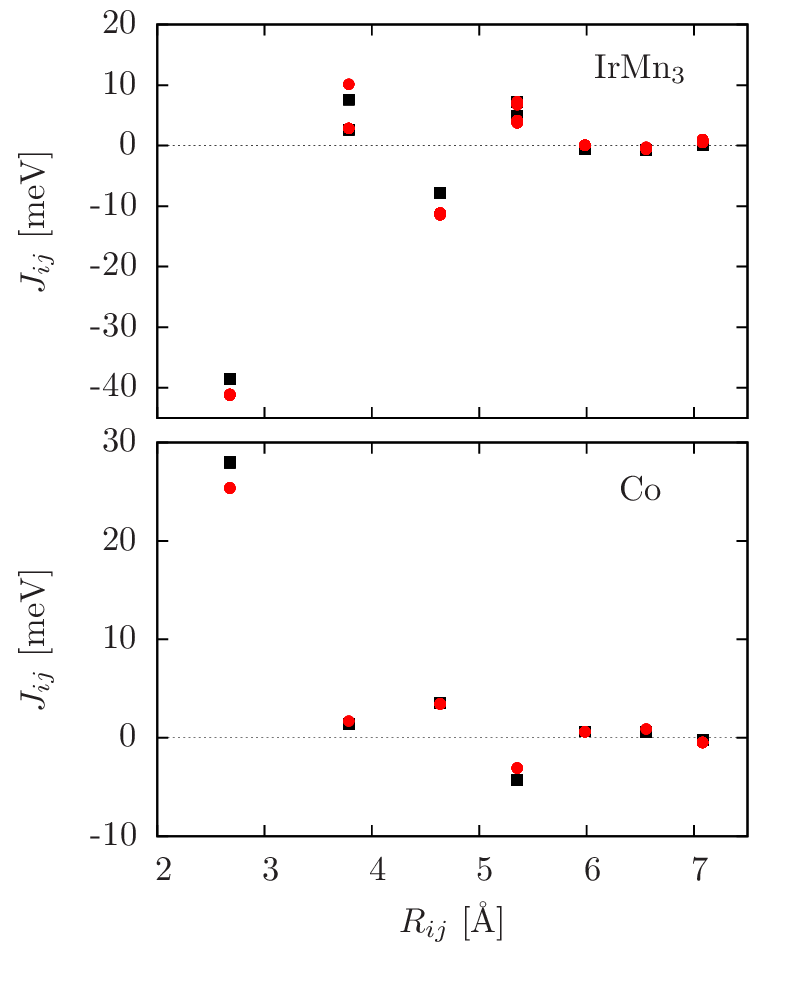} 
\caption{(Colour online) Calculated isotropic exchange interactions, $J_{ij}$, for bulk L1$_{2}$ IrMn$_3$ (upper panel) and fcc bulk Co (lower panel).  Squares and circles refer to the  $J_{ij}$'s obtained from the DLM states by using the SCE-RDLM method and by using the RTM,\cite{Rtorque1} using the ordered ground state (the $T$1 state for IrMn$_3$ and the ferromagnetic state for Co) respectively.  The interactions are displayed as a function of the distance, $R_{ij}$.}
\label{fig:bulk-jij}
\end{figure}

Recently, we investigated the magnetic ground-states of the ordered IrMn$_3$ alloy.~\cite{IrMn-MAE}  In agreement with experiment~\cite{tomenoJAP99} and other theoretical work~\cite{sakumaPRB03}, we found that the ground-state spin-structure of the Mn sublattices is a frustrated 120$^\circ$ N\'eel-type state in the (111) planes of the fcc lattice, called the $T1$ state.  This frustration is the consequence of the strong antiferromagnetic exchange coupling between the Mn atoms.  In Fig.~\ref{fig:bulk-jij} we compare the isotropic exchange interactions,  $J_{ij}=(1/3) \,\mbox{Tr} \, {\bf J}_{ij}$, obtained from the newly developed SCE-RDLM scheme with those from the RTM.\cite{Rtorque1}  Clearly the two sets of parameters (circles and squares) are in very good quantitative agreement.  This indicates that the softening of the magnetic moments in the DLM state do not imply a softening of the exchange couplings with respect to the ordered state.  In contrast to IrMn$_3$, the exchange interactions of bulk Co are primarily of ferromagnetic nature.  A very good agreement between the SCE-RDLM and the RTM can also be seen in this case.  Importantly, both systems are characterized by short-range interactions: the $J_{ij}$'s practically vanish beyond the fourth neighbour shell.

For the 6ML IrMn$_3$/12ML Co/6ML IrMn$_3$ interface system we considered all the pairs of magnetic atoms separated by a distance less than $10.68$\AA.  This implied the calculation of a total number of more than 13,000 tensorial interactions.  
Inspection of the calculated isotropic exchange interactions reveals that the Mn-Mn and Co-Co interactions are spread around their corresponding bulk values due to symmetry breaking at the interface: the nearest neighbor (NN) Mn-Mn interactions are somewhat increased while most of the NN Co-Co interactions are decreased.  This decrease is of up to 20\% near the interface.  This feature clearly correlates with the change of the local magnetic moments caused by the presence of the interface depicted in Fig.~\ref{fig:moments}.  The Mn-Co interactions across the interface are quite small in magnitude ($<$ 10~meV) and mostly antiferromagnetic: $J_{ij}=-6.15 \,$meV for nearest neighbours, while  $J_{ij}=-4.86 \,$meV and $-8.40$~meV for next nearest neighbours. 

In order to visualize the isotropic exchange interactions for the interface system, we defined the following effective interaction parameters,
\begin{equation}
J_{pi} = \sum_{q,j} J_{pi,qj} \, \vec{e}_{pi} \cdot \vec{e}_{qj} \; ,
\label{eq:Jpi}
\end{equation}
where $p$ and $q$ denote layers, while $i$ and $j$ stand for sites within the corresponding layers.  For this calculation we choose a magnetic state, $\{ \vec{e}_{pi} \}$, corresponding to the $T$1 structure for the IrMn$_3$ side and to an FM state perpendicular to the interface within the Co layers.  This configuration was chosen because of the antiferromagnetic nature of the interface coupling combined with the large Mn anisotropy, which constrains the $T$1 structure within the interface plane. 
\begin{figure}[htb]
\includegraphics{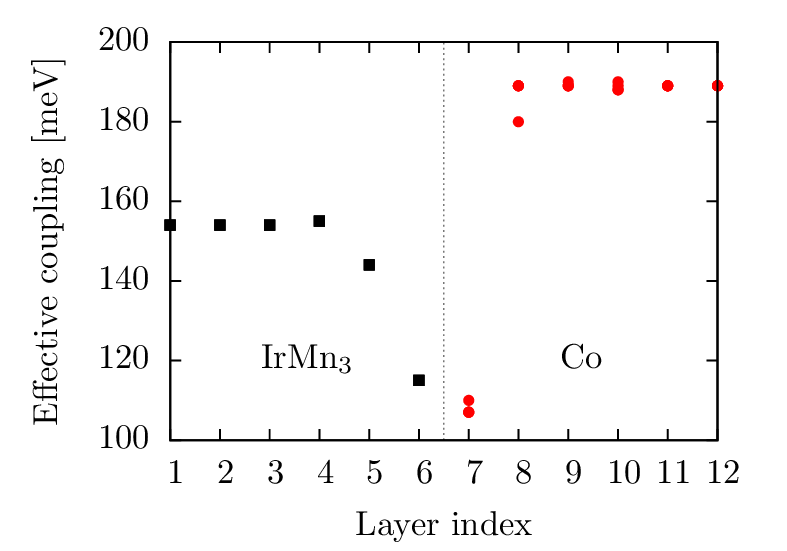}
\caption{Calculated effective exchange interactions, $J_{pi}$, for the IrMn$_3$/Co interface as defined in Eq.~(\ref{eq:Jpi}).  The labelling of layers is identical to that in Fig.~\ref{fig:moments}.}
\label{fig:interface-jij}
\end{figure}
Clearly, by translational and rotational symmetry of this spin-structure,  $J_{pi}$ turned out to be the same for all the Mn sites in a given layer $p$ of the IrMn$_3$ region, while in each layer of the Co region, in principle, two different values for $J_{pi}$ may be obtained.  In particular for the Co layer adjacent to IrMn$_3$, these two different values are associated with Co sites having two and three NN Mn atoms. 

Fig.~\ref{fig:interface-jij} clearly shows that the effective exchange interactions quickly approach their bulk value with distance from the interface.  This follows from the fact that only interactions between atoms in close proximity to the interface deviate from their counterparts in the bulk; those more than 2 atomic planes away from the interface are practically bulk-like.  Such behavior is to be expected because of the very narrow range of exchange interactions in the bulk systems.  Since for the chosen magnetic states the magnetic orientations of the Mn and Co atoms are perpendicular to each other, the otherwise weak Mn-Co interactions are excluded from the sum in Eq.~(\ref{eq:Jpi}) and the the effective interactions are significantly decreased near the interface.  The different values of $J_{pi}$  for the non-equivalent Co sites near to the interface can also be seen in the figure.

\subsection{Magnetic anisotropies}
The most remarkable observation of our previous study~\cite{IrMn-MAE} was the surprisingly strong, second order MA of IrMn$_3$ resulting from the fact that the cubic symmetry is locally broken for each of the three Mn sub-lattices.  Rotating the $T1$ state around the axis normal to the $(111)$ plane, the energy of the system can be described as $E(\varphi)=E_0 + K_{\mathrm{eff}} \sin^2 \varphi$.  By calculating $E(\varphi)$ in the $T1$ spin-state, a value of $K_{\mathrm{eff}}$=10.4 meV was deduced.~\cite{IrMn-MAE}  We repeated this calculation by using the self-consistent field potentials from the DLM state and obtained a value of $K_{\mathrm{eff}}=7.67 \,$meV.  Although this value is about 30\% less than the one we obtained in our previous calculation using the $T1$ ground-state, it is still consistent with the observation of a very large magnetic anisotropy.

The SCE-RDLM scheme provides a unique opportunity to calculate contributions of the MAE related to the one-site and the two-site exchange anisotropy terms, which is an important issue for spin-model simulations because it affects the scaling of the anisotropy with magnetisation.  First we note that, by symmetry, in the frame of reference with $z$ axis along the (111) direction, the one-site anisotropy matrices of the three Mn sublattices in the L1$_2$ phase can be expressed as,
\begin{eqnarray}
\mathbf{K}_{1}^{\left(  111\right)  } &=& K\left(
\begin{array}[c]{ccc}
0 & 0 & 0\\
0 & \frac{2}{3} & -\frac{\sqrt{2}}{3}\\
0 & -\frac{\sqrt{2}}{3} & \frac{1}{3}%
\end{array} \right)  \; ,  \\ \nonumber \label{eq:K1} \\
\mathbf{K}_{2}^{\left(  111\right)  } &=& K\left(
\begin{array}[c]{ccc}
\frac{1}{2} & -\frac{\sqrt{3}}{6} & -\frac{\sqrt{6}}{6}\\
-\frac{\sqrt{3}}{6} & \frac{1}{6} & \frac{\sqrt{2}}{6}\\
-\frac{\sqrt{6}}{6} & \frac{\sqrt{2}}{6} & \frac{1}{3}%
\end{array} \right)  \; ,  \\ \nonumber  \label{eq:K2} \\
\quad\mathbf{K}_{3}^{\left(  111\right)  } &=& K\left(
\begin{array}[c]{ccc}
\frac{1}{2} & \frac{\sqrt{3}}{6} & \frac{\sqrt{6}}{6}\\
\frac{\sqrt{3}}{6} & \frac{1}{6} & \frac{\sqrt{2}}{6}\\
\frac{\sqrt{6}}{6} & \frac{\sqrt{2}}{6} & \frac{1}{3}%
\end{array}
\right) \; .
\label{eq:K3}
\end{eqnarray}
Our calculations based on Eqs.~(\ref{eKe}) and (\ref{Jone-b}) confirmed with high accuracy the above non-trivial forms of the one-site anisotropy matrices, with a value of $K$ = 0.82 meV.  Note that the total one-site contribution to $K_{\mathrm{eff}}$ is $2K$.\cite{IrMn-MAE}  In addition, from the tensorial exchange interactions, Eqs.~(\ref{Jtwo}) and (\ref{eJe}), we calculated a two-site anisotropy contribution of 6.73 meV to the MAE.  Thus, the spin-Hamiltonian derived from the SCE-RDLM approach yielded $K_{\mathrm{eff}}$ = 8.37 meV, in good agreement with $K_{\mathrm{eff}}$ = 7.67 meV we calculated directly from the band-energy.  The difference between the two values should be related to the fact that the direct calculation of the MAE is not restricted to second order interactions as is the case for the spin-Hamiltonian.  Due to cubic symmetry, for fcc bulk Co we calculated negligible MAE, $|K_{\mathrm{eff}}|< 1 \, \mu$eV.

Next we investigated the magnetic anisotropy of the spin-structure described in the context of Eq.~(\ref{eq:Jpi}) and Fig.~\ref{fig:interface-jij} for the interface system.  Of particular interest is the variation of the energy, Eq.~(\ref{extHeis}), during a magnetization reversal process.  For transparency, we considered the reversal of the Co moments from $z$ to $-z$ by keeping the relative orientations of any two moments in the system fixed.  Theoretically, this is equivalent to rotating all the spins simultaneously around an axis in the (111) plane by an angle of $0 \le \varphi \le \pi$.  Clearly from Eq.~(\ref{extHeis}) the energy, $E(\varphi)$, can be recast into layer-wise contributions, $E_p(\varphi)$, such that
\begin{equation}
E(\varphi)= N \sum_p E_p(\varphi) \; , 
\label{eq:anis}
\end{equation}
where $N$ is the number of cells in a layer and $E_p(\varphi)$ comprises the contributions of three Mn atoms and four Co atoms in the IrMn$_3$ and Co parts of the interface system.  Moreover, $E_p(\varphi)$ can be decomposed into contributions arising from the one- and two-site terms of Eq.~(\ref{extHeis}).  
\begin{figure}[htb]
\includegraphics{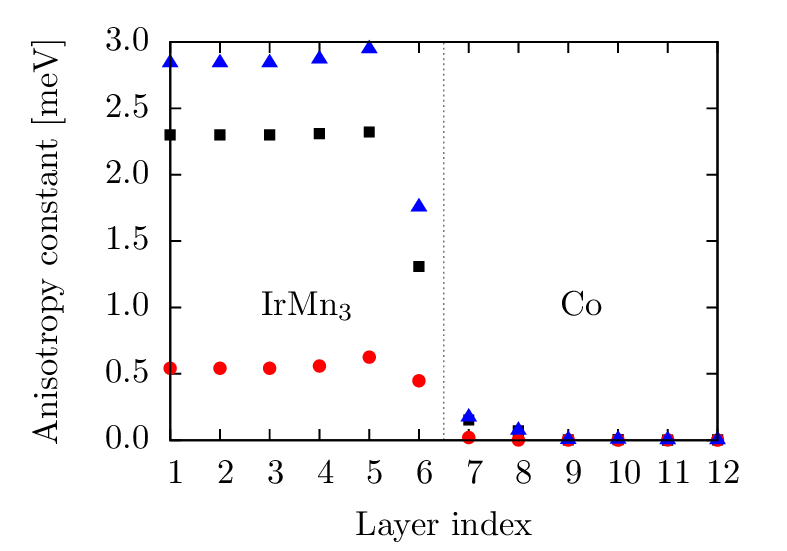}
\caption{(Colour online) Calculated effective anisotropy parameters (triangles), $K^p_{\mathrm{eff}}$, see Eq.~(\ref{eq:anis}), for the IrMn$_3$/Co interface.  Also shown are the corresponding one-site ($K^{p,\mathrm{os}}_{\mathrm{eff}}$, circles) and two-site ($K^{p,\mathrm{ts}}_{\mathrm{eff}}$, squares) contributions.}
\label{fig:anisotropy}
\end{figure}

Note that in these calculations only the  symmetric part of the tensorial exchange matrices,
${\bf J}_{ij}^s=\frac{1}{2}\left({\bf J}_{ij}+{\bf J}_{ij}^T\right)$,
 were considered.  Our numerical results confirmed that the following forms of $E_p(\varphi)$, which describe the bulk anisotropy energies,\cite{IrMn-MAE} give a good description for the layered system as well,
\begin{eqnarray}
E_p(\varphi) &=& E_p(0) + \frac{K^p_{\mathrm{eff}}}{8} \Big( 2 + \sin^2\varphi
-2 \cos\varphi \nonumber \\
&& -2\sqrt{2} \sin\varphi (1-\cos\varphi) \Big) \; ,
\label{eq:EL12B}
\end{eqnarray}
for the IrMn$_3$ layers and
\begin{equation}
E_p(\varphi)=E_p(0) + K^p_{\mathrm{eff}} \sin^2 \varphi \; ,
\end{equation}
for the Co layers. 

In Fig.~\ref{fig:anisotropy} the layer-resolved effective anisotropy parameters, $K^p_{\mathrm{eff}}$, are normalized to one Mn (layers no. 1--6) and Co (layers no. 7--12) atom, together with their one-site and two-site contributions.  Reassuringly, when departing from the interface these functions show a stable convergence to the corresponding bulk values, $K^p_{\mathrm{eff}}$=2.79 meV, $K^{p,\mathrm{os}}_{\mathrm{eff}}$=0.54 meV and $K^{p,\mathrm{ts}}_{\mathrm{eff}}$= 2.28 meV for IrMn$_3$, and $K^p_{\mathrm{eff}}$  $\simeq$ $K^{p,\mathrm{os}}_{\mathrm{eff}}$ $\simeq$ $K^{p,\mathrm{ts}}_{\mathrm{eff}}$ $\simeq 0$  for Co.  In the vicinity of the interface, $K^{p,\mathrm{os}}_{\mathrm{eff}}$ of Mn shows moderate oscillations and $K^{p,\mathrm{os}}_{\mathrm{eff}}$ of Co is still below the 0.01--0.02 meV range.   The two-site anisotropy contribution at the IrMn$_3$ layer nearest to the interface drops, however, to half of the bulk value.  The absence of Mn-Mn interactions at the interface reduces the nearest neighbour coordination of the interface Mn spins from 8 to 6, the decrease observed indicates therefore that some softening of the intrinsic Mn anisotropy is found at the interface.  Breaking of the cubic bulk symmetry, i.e., missing of the Co planes from one side, also results in an increase to about 0.22 meV of the Co $K^{p,\mathrm{ts}}_{\mathrm{eff}}$ at the interface.

\subsection{Dzyaloshinskii-Moriya interactions}
The antisymmetric components of the exchange interactions, ${\bf J}_{ij}^a=\frac{1}{2}\left({\bf J}_{ij}-{\bf J}_{ij}^T\right)$, correspond to the well known Dzyaloshinskii-Moriya (DM) interaction.\cite{DM1,DM2}  The DM interaction favours perpendicular alignment of moments and for a given pair of spins, the antisymmetric component of the exchange interaction may be conveniently represented by the 
DM vector, $\vec{D}_{ij}$,
\begin{equation}
\vec{e}_i {\bf J}_{ij}^a \vec{e}_j = \vec{D}_{ij}\cdot \left( \vec{e}_i\times\vec{e}_j  \right)\; .
\label{eq:DMvec}
\end{equation}
In the L1$_2$ structure, the absence of inversion symmetry at the Mn lattice points gives rise to a DM interaction with a magnitude of 1.24 meV between nearest neighbours, with $\vec{D}_{ij}$ aligned along the ($1\bar{2}\bar{2}$) directions.  
Following the trend of the isotropic exchange interactions, the magnitudes of the DM interactions diminish very rapidly with distance; the third and fifth NN interactions in the bulk are 0.58 and 0.13 meV respectively.  
Clearly enough, in a non-collinear ground state of a bulk system, the asymmetric contribution to the exchange interaction 
might be non-zero only for different sublattices. However, due to symmetry, 
the inter-sublattice DM interactions also give rise to no overall 
magnetic anisotropy energy in bulk L1$_2$ IrMn$_3$.

Calculations reveal that the symmetry breaking at the IrMn$_3$/Co interface gives rise to additional asymmetry in the exchange interactions at the interface.  Although restricted to the atomic planes in the immediate vicinity of the interface, these interactions cause an anisotropy contribution for the reversal process discussed in the context of Fig. \ref{fig:anisotropy} that arises purely due to the presence of the interface and shows an angular dependence that is unidirectional,
\begin{equation}
E_p^{\rm DM}(\varphi)=E_p(0) + K_p^{\rm DM} \cos \varphi \; .
\label{eq:DMvarphi}
\end{equation}

\begin{figure}[htb]
\includegraphics{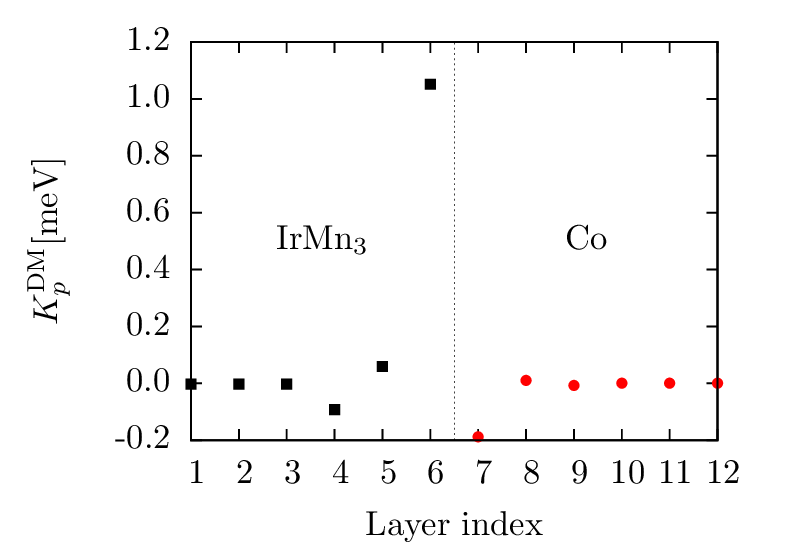}
\caption{(Colour online) Calculated layer-wise anisotropy constants due to DM interactions, $K_p^{DM}$, see Eq.~(\ref{eq:DMvarphi}), for the IrMn$_3$/Co interface.}
\label{fig:DMenergy}
\end{figure}

$K_p^{\rm DM}$ is plotted for the interface structure in Fig. \ref{fig:DMenergy}.  The reduced coordination of the interface Mn spins gives rise to a large anisotropic DM contribution to the system energy.  Specifically, at the interface the antisymmetric exchange components give rise to an additional energy difference between the $T$1 and $T$2 Mn spin configurations, related to different
chirality.\cite{IrMn-MAE}  

\begin{figure}[htb]
\includegraphics{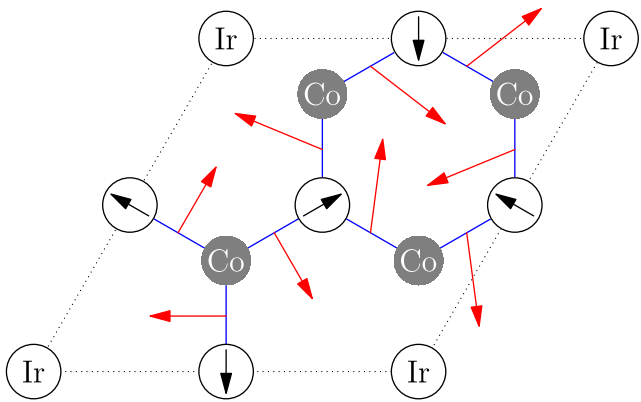}
\caption{(Colour online) The arrows show the (111) in-plane projections of the DM vectors, $\vec{D}_{ij}$, for nearest neighbour interface Co-Mn interactions.  The $T$1 configuration of the Mn spins is indicated by arrows within circles at the Mn positions. Note that the plane of the Co atoms are above the IrMn$_3$ layer. }
\label{fig:dm-Interface}
\end{figure}

Concerning the Co layers, a value of $-0.19$ meV for the DM anisotropy energies (shown in Fig. \ref{fig:DMenergy}) denotes a unidirectional anisotropy favoring alignment of the Co spins perpendicular to, or away from, the interface plane.  This anisotropy arises in the antisymmetric components of the interface Co-Mn interactions.  The DM vectors for the interface exchange interactions are depicted as solid arrows in Fig. {\ref{fig:dm-Interface}}, where $\vec{D}_{ij}$ is shown when $i$ indexes a Co site and $j$ a Mn site.

The interface DM vectors are found to lie almost entirely in the interface plane which favours a perpendicular spin alignment of the Co relative to the interface.  Which of the two perpendicular directions is favoured depends upon the sense of the DM vectors and the Mn spin configuration: for the Mn configuration shown in Fig. {\ref{fig:dm-Interface}}, Co alignment away from the IrMn$_3$ layer is preferred.  
Simultaneous inversion of the spins on all three Mn sublattices, which does not affect the chirality of the Mn spin structure, produces a configuration that is energetically equivalent in the bulk, but which causes the preferred orientation of the Co layers at an interface to be reversed.  Note, however, that such a change in the Mn spin configuration cannot be obtained by any single global rotation.  

\section{Conclusions}
We have introduced a new computational scheme to calculate magnetic anisotropy parameters and tensorial
exchange interactions based on a Spin-Cluster Expansion and the Relativistic Disordered Local Moment picture of
itinerant magnets. The method relies on the high-temperature paramagnetic phase, thus, it is,
uniquely applicable to  any system with arbitrary magnetic ground state.  In particular, it is highly efficient in
cases when the complex magnetic structure of the system is not known a priori from first principles
total energy calculations, such as for heterogeneous interfaces and nanoparticles. Moreover, the SCE-RDLM method
can be used to calculate higher order multispin interactions.  The main limitation of the method is that 
longitudinal spin-fluctuations are neglected: in case of large induced moments further extension
of the theory is needed.\cite{Mryasov05} 

We applied the new method to calculate magnetic anisotropies and spin-interactions at an IrMn$_3$/Co(111) interface. 
Our main observation is that the corresponding bulk properties are influenced in the immediate vicinity of the
interface only. The exchange interaction between the Mn and Co atoms are weak leaving the corresponding
bulk magnetic structures fairly unaffected even at the interface. Symmetry breaking at the interface gives,
however, rise to large anisotropy effects. Most prominently, the appearance of sizeable Dzyaloshinskii-Moriya 
interactions at the interface results in to a perpendicular coupling of the Co and Mn spins and to unidirectional exchange anisotropy
that might substantially influence the exchange bias in such systems.\cite{Thomas98,Krivorotov03,Dong09} By using the
spin-model parameters derived in this work,  we are going to present a detailed study to explore the spin-dynamics and the exchange 
bias properties of the IrMn$_3$/Co(111) interface.
    
\acknowledgments
Financial support was provided by the Hungarian Research Foundation (contract no. OTKA K68312 and K77771)
and by the New Hungary Development Plan (Project ID: T\'AMOP-4.2.1/B-09/1/KMR-2010-0002).  
J.J. acknowledges financial support by Seagate Technology, Northern Ireland.

\appendix
\section{Non-relativistic limit of the tensorial exchange interactions}

In order to derive a non-relativistic limit of the SCE-RDLM formalism for the two-site exchange interactions we first expand
the logarithm to first order in Eq. (\ref{Jtwo-b}), 
\begin{equation}
J^{LL'}_{ij} = \frac{1}{\pi} \operatorname{Im}\int^{\varepsilon_F} d\varepsilon
\, \mathrm{Tr} \left(
\underline{X}_{i,L}  \, \underline{\tau}_{c,ij} \,
\underline{X}_{j,L'} \,  \underline{\tau}_{c,ji}  \right)
\;,
\label{Jtwo-b1}
\end{equation}
with
\begin{equation}
\underline{X}_{i,L} =
\int d^2 e_{i}  \,
\underline{X}_i\left(\vec{e}_i\right)
\, Y_L^\ast(\vec{e}_i)
\;,
\label{XiL}
\end{equation}
where, for convenience, but in this section only, we use complex spherical harmonics.  In the non-relativistic case the transformation (\ref{t-e}) applies in spin space only,
\begin{equation}
\underline{t}_{i}^{ss^{\prime}}\left(  \vec{e}_{i}\right)  =
  \underline{t}_{i,0} \, \delta_{ss^{\prime}} + \Delta\underline{t}_{i}\,
  \vec{\sigma}_{ss^{\prime}} \, \vec{e}_{i}\;,
\label{t-e-nr}
\end{equation}
where $\underline{t}_{i,0}$ and $\Delta\underline{t}_{i}$ are now matrices in non-relativistic angular momentum $(\ell,m)$-space, $\vec{\sigma}$ are the Pauli matrices and $s=\uparrow,\downarrow$ denotes the spin-index.  Clearly, the usual spin-up and spin-down components of the {\it t}-matrices are,
\begin{align}
\underline{t}_{i}^{\uparrow}  &  =\underline{t}_{i,0}+\Delta\underline{t}%
_{i}\;, \label{t-u} \\
\underline{t}_{i}^{\downarrow}  &  =\underline{t}_{i,0}-\Delta\underline{t}%
_{i}\;,
\label{t-d}
\end{align}
while for the paramagnetic case the effective {\it t}-matrices and SPO matrices become diagonal in spin space,
\begin{eqnarray}
\underline{t}_{i,c}^{ss^{\prime}} & = & \underline{t}_{i,c} \, \delta_{ss^{\prime}} \;,
\label{tc-nr}  \\
\underline{\tau}_{ij,c}^{ss^{\prime}} & = & \underline{\tau}_{ij,c} \, \delta_{ss^{\prime}} \;.
\label{tauc-nr}
\end{eqnarray}
Since also the excess scattering matrices take the form of (\ref{t-e-nr}), (\ref{t-u}) and
(\ref{t-d})
\begin{equation}
\underline{X}_{i}^{ss^{\prime}}\left(  \vec{e}_{i}\right)  = \underline{X}_{i,0} \, \delta_{ss^{\prime}}
+ \Delta\underline{X}_{i}\,\overrightarrow
{\sigma}_{ss^{\prime}} \, \vec{e}_{i}\;,
\end{equation}
with
\begin{align}
\underline{X}_{i}^{\uparrow}  &  =\underline{X}_{i0}+\Delta\underline{X}%
_{i}\;,\\
\underline{X}_{i}^{\downarrow}  &  =\underline{X}_{i0}-\Delta\underline{X}%
_{i}\;,
\end{align}
and
\begin{equation}
\underline{X}_{i}^{s}=\left[  \left(  \underline{t}_{i,c}^{-1}-\left( \underline{t}%
_{i}^{s}\right)^{-1} \right)^{-1}-\underline{\tau}_{ii,c}\right]  ^{-1}\qquad\left(
s=\uparrow,\downarrow\right)  \;,
\end{equation}
the CPA condition (\ref{CPA-1}) can be rewritten as,
\begin{equation}
\frac{1}{2} \left( \underline{X}_{i}^{\uparrow} + \underline{X}_{i}^{\downarrow} \right) = 0    \;.
\label{CPA-nr}
\end{equation}

The integration over angular variables can be explicitly performed in Eq. (\ref{XiL}),
\begin{align}
& \underline{X}_{i,L}  = \Delta\underline{X}_{i} \: \vec{\sigma}
\int d^{2} {e}_{i} \, Y_{L}^{\ast} \, \left(
\vec{e}_{i}\right)  \, \vec{e}_{i} \label{XiL2}  \\
& \left(\frac{4\pi}{3}\right)^{\frac{1}{2}} \Delta\underline{X}_{i} 
\left(  2\delta_{L,\left(
1,-1\right)  } \, \sigma_{+}-2\delta_{L,\left(  1,1\right)  } \, \sigma_{-}%
+\delta_{L,\left(  1,0\right)  } \, \sigma_{z}\right)  \;,
\nonumber
\end{align}
with $\sigma_\pm=\sigma_x \pm i \sigma_y$.  The trace in Eq. (\ref{Jtwo-b1}) can then be recast into a trace in angular momentum space, $Tr_{L}$, and a trace in spin space, $ \mathrm{Tr}_{s}$,
\begin{align}
&   \mathrm{Tr} \left(  \underline{X}_{i,L} \, \underline{\tau}_{c,ij} \, \underline
{X}_{j,L^{\prime}} \, \underline{\tau}_{c,ji} \right)
\nonumber \\
& =\frac{4\pi}{3} \,  \mathrm{Tr}_{L}\left(  \Delta\underline{X}_{i} \, \underline{\tau
}_{c,ij} \, \Delta\underline{X}_{j} \, \underline{\tau}_{c,ji} \right)
\nonumber \\
&   \qquad \quad \mathrm{Tr}_{s} \left[  \left(  2\delta_{L,\left(  1,-1\right)} \, \sigma_{+}%
-2\delta_{L,\left(  1,1\right)} \, \sigma_{-}+\delta_{L,\left(  1,0\right)}
\, \sigma_{0}\right)  \right. 
\nonumber \\
&  \qquad  \qquad  \left. \left(  2\delta_{L^{\prime},\left(  1,-1\right)} \,
\sigma_{+}-2\delta_{L^{\prime},\left(  1,1\right)} \, \sigma_{-}+\delta
_{L^{\prime},\left(  1,0\right)} \, \sigma_{0}\right)  \right]  \;.
\end{align}

By using the algebra of the Pauli matrices, the trace in spin space can easily be performed and the two-site terms in the spin-Hamiltonian will be transformed into a pure isotropic Heisenberg interaction,
\begin{equation}
\sum_{LL^{\prime}}J_{ij}^{LL^{\prime}} \, Y_{L}\left(  \vec{e}_{i}\right)
Y^*_{L^{\prime}}\left(  \vec{e}_{j}\right)  = J_{ij}\,\vec{e}_{i}%
\vec{e}_{j}\;,
\end{equation}
with
\begin{align}
J_{ij} & =-\frac{2}{\pi}\operatorname{Im}\int^{\varepsilon_F} d\varepsilon\,
 \mathrm{Tr}_{L}\left(  \Delta\underline{X}_{i} \, \underline{\tau
}_{c,ij} \, \Delta\underline{X}_{j} \, \underline{\tau}_{c,ji} \right)
\\ &
=-\frac{1}{2\pi}\operatorname{Im}\int^{\varepsilon_F} d\varepsilon\,  \mathrm{Tr}_{L}\left[
\left( \underline{X}_{i}^\uparrow - \underline{X}_{i}^\downarrow \right)
\underline{\tau }_{c,ij} \right. 
\nonumber \\ &
\qquad \qquad \qquad \qquad \qquad  
\left. \left( \underline{X}_{j}^\uparrow - \underline{X}_{j}^\downarrow \right)
\underline{\tau}_{c,ji} \right]
\;.
\end{align}
The above formula is widely used in non-relativistic calculations of the exchange interactions.

\bibliographystyle{myst}

\end{document}